\documentclass[11pt,letterpaper]{article}
\usepackage[margin=1in]{geometry}
\usepackage[T1]{fontenc}
\usepackage{lmodern,amsmath,amssymb,amsthm,microtype}
\usepackage[hidelinks]{hyperref}
\usepackage{enumitem}
\usepackage{tikz,needspace}
\usetikzlibrary{shapes.geometric}
\definecolor{bpblue}{RGB}{39,91,137}
\definecolor{bpred}{RGB}{171,65,50}
\definecolor{bpgray}{RGB}{100,108,116}
\definecolor{bpsp}{RGB}{137,185,224}
\definecolor{bpsm}{RGB}{211,231,247}
\definecolor{bpx}{RGB}{239,166,157}
\definecolor{bpf}{RGB}{218,222,227}
\definecolor{bpemone}{RGB}{126,193,165}
\definecolor{bpemtwo}{RGB}{211,235,211}
\definecolor{bpepone}{RGB}{240,181,104}
\definecolor{bpeptwo}{RGB}{250,226,166}
\tikzset{
	bpedge/.style={draw=bpgray,line width=.7pt},
	bpbin/.style={draw=bpgray,line width=.7pt,rounded corners=2pt},
	bpitem/.style={rectangle,rounded corners=1pt,draw=black!30,
		line width=.4pt,minimum height=7mm,inner sep=1pt},
	bpsp/.style={fill=bpsp},bpsm/.style={fill=bpsm},
	bpx/.style={fill=bpx},bpf/.style={fill=bpf},
	bpemone/.style={fill=bpemone},bpemtwo/.style={fill=bpemtwo},
	bpepone/.style={fill=bpepone},bpeptwo/.style={fill=bpeptwo},
	bpvertex/.style={circle,fill=bpgray,draw=white,line width=.3pt,
		inner sep=0pt,minimum size=3pt},
	bpredmark/.style={regular polygon,regular polygon sides=3,
		fill=bpred,draw=bpred,inner sep=0pt,minimum size=5pt},
	bpbluemark/.style={rectangle,fill=bpblue,draw=bpblue,
		inner sep=0pt,minimum size=4pt}
}
\setlist[enumerate]{label=\textup{(\roman*)},leftmargin=*,itemsep=3pt,topsep=5pt}
\newtheorem{theorem}{Theorem}[section]
\newtheorem{lemma}[theorem]{Lemma}
\newtheorem{proposition}[theorem]{Proposition}

\newtheorem{claim}{Claim}[section]
\newenvironment{claimproof}{\begin{proof}[Proof of the claim]}{\end{proof}}
\theoremstyle{definition}

\numberwithin{equation}{section}

\newcommand{\OPT}{\operatorname{OPT}}
\newcommand{\LP}{\operatorname{LP}}

\newcommand{\ls}{\log^{*}}
\newcommand{\Q}{\mathbb Q}
\newcommand{\Z}{\mathbb Z}
\newcommand{\OAIurl}{https://github.com/openai/math/blob/main/preprints/Additive-hardness-and-unbounded-configuration-gaps-in-bin-packing-September-24-2026/Additive-hardness-and-unbounded-configuration-gaps-in-bin-packing-September-24-2026.pdf}
\hypersetup{pdftitle={Additive Hardness and Configuration-LP Gaps for Three-Item Bin Packing},pdfauthor={Ilan Doron-Arad}}
\begin{document}
	\hypersetup{pageanchor=false}
	\begin{titlepage}
		\centering
		\vspace*{0.6in}
		{\LARGE Additive $\Omega\left(\log^* \OPT\right)$ Gap for Bin Packing\\
			with Three Items per~Bin\par}
		\vspace{0.4in}
		{\large Ilan Doron-Arad\par}
		\vspace{5pt}
		MIT\par
		\vspace{0.15in}
		\vspace{0.5in}
		\begin{abstract}\normalsize
			We prove that bin packing is NP-hard to approximate within an additive factor of
			$o(\ls\OPT)$, even when all item sizes are arbitrarily close to $1/3$. This gives the first lower bound for bin packing with an additive factor that depends explicitly on the optimum value.
		 In addition, we give a polynomial-time reduction
from any bin packing instance into one with item
sizes arbitrarily close to $1/3$, preserving its additive
configuration-LP gap up to constant factors.
Thus, in the sense of additive configuration LP gap and up to a constant,
we confirm the conjecture that 3-partition instances are the hardest bin packing instances.
In particular, we obtain an unconditional $\Omega(\ls\OPT)$
gap for 3-partition instances. We build on the recent OpenAI construction that uses five items per bin and rules out any
constant additive gap.
		\end{abstract}
	\end{titlepage}
	\setcounter{page}{1}
	\hypersetup{pageanchor=true}

	\section{Introduction}
	In the {\em bin packing} problem, we are given a set of items $I$, each item $i\in I$ has a size $s_I(i)\in(0,1]$. A {\em packing} is a partition of $I$ into subsets each of total size at
	most one. Its cost is the number of subsets, called {\em bins}; let $\OPT(I)$
	be the minimum number of bins for instance $I$. 
	
	Bin packing is one of the classical problems in combinatorial optimization. Even in the restricted setting of \textsc{3-Partition}, where all item sizes lie in $(1/4,1/2)$, the problem is strongly NP-hard~\cite{GJ79}. As a result, a long line of research has developed efficient approximation algorithms for the problem. Among the earliest are simple greedy heuristics: Next Fit uses at most $2\OPT-1$ bins,  while First Fit Decreasing (FFD), which processes items in nonincreasing order of size, achieves the factor $\frac{11}{9}\OPT+\frac{2}{3}$~\cite{Joh73,Dos07}. 
	
	A major breakthrough was the {\em asymptotic polynomial time approximation scheme} of Fernandez de la Vega and Lueker~\cite{VL81}, which obtains $(1+\varepsilon)\OPT+O_\varepsilon(1)$ bins for every fixed $\varepsilon>0$ with running time $n^{O_{\varepsilon}(1)}$. Next, further algorithmic progress was done to remove the multiplicative $(1+\varepsilon)$ factor, leaving only an {\em additive} term. Namely, the seminal work of Karmarkar and Karp~\cite{KK82} developed an algorithm, using at most $\OPT+O(\log^2\OPT)$ bins, building on the configuration LP introduced by Gilmore and Gomory~\cite{GG61}. This linear programming relaxation assigns fractional weights to feasible bin configurations and has played a central role in subsequent approximation algorithms. Rothvoss~\cite{Rot13} improved the additive guarantee to $O(\log\OPT\log\log\OPT)$, and Hoberg and Rothvoss~\cite{HR17} subsequently improved it to $O(\log\OPT)$, which remains the best known additive approximation guarantee. 
	
	The additive integrality gap of the configuration LP itself has been studied directly. A longstanding conjecture, known as the {\em Modified Integer Round-Up Property} (MIRUP)~\cite{ST97}, asserted that $\OPT\le\lceil\LP\rceil+1$, even for generalizations of bin packing. More generally, whether the configuration LP has a constant additive integrality gap was listed by Williamson and Shmoys~\cite{WS11} among ten major open problems in approximation algorithms. For the three-item case (3-partition), Eisenbrand, Pálvölgyi, and Rothvoss~\cite{EPR13} established a connection to the discrepancy of three permutations, showing that a constant discrepancy bound would imply a constant additive integrality gap.
	
	Despite this progress on upper bounds, additive lower bounds of more than $2$ remained elusive for a long time. Recently, OpenAI~\cite{OAI26} proved NP-hardness for every constant additive factor and constructed instances with arbitrarily large constant additive gaps for the configuration LP. Their construction uses item sizes greater than $1/6$, so each bin contains at most five items. In this paper, we strengthen both the computational hardness and the integrality gap. We do so even when all item sizes are arbitrarily close to $1/3$, so each bin contains at most three items. Specifically, we establish $\Omega(\log^*\OPT)$ additive NP-hardness and construct instances whose configuration LP has an additive integrality gap of $\Omega(\log^*\OPT)$ and show that any stronger additive gap can be obtained, up to a constant factor, in this family of instances where all items are arbitrarily close to $1/3$. 
	
	\subsection{Definitions}
We define the configuration LP and the conventions used to state our results.
	Throughout, assume that
	inputs list all items explicitly, and running
	times are measured in the binary input length. For a function $f$ on
	an item set and a subset $U$, write $f(U)=\sum_{i\in U}f(i)$.
	For $x\ge1$, let $\ls x$ be the number of base two logarithms needed to be applied iteratively
	to obtain a value at most one.
	For an instance $I$ of bin packing, define its configurations by
	\[
	\mathcal H(I)=\{U\subseteq I:s_I(U)\le1\}.
	\]
	The corresponding LP is
	\[
	\LP(I)=\min\left\{\sum_{U\in\mathcal H(I)}y_U:
	y_U\ge0\ \forall U\in\mathcal H(I),\quad \textnormal{s.t. }
	\sum_{U\ni i}y_U\ge1\ \forall i\in I\right\}.
	\]
	
	\subsection{Results}
	Our main result is a lower bound of $\Omega\left(\ls\OPT\right)$ for bin packing with at most three items per bin. 
	
	\begin{theorem}\label{thm:hardness}
		There is an absolute constant $\alpha>0$ such that, for every fixed $\varepsilon>0$, the following promise problem is $\textnormal{NP-hard}$.
		An input is a bin packing instance $J$ of $3M$ items for an integer $M\ge1$, with sizes in $
		(1/3-\varepsilon,1/3+\varepsilon)\cap(1/4,1/2)$. 
		The two cases are
		\begin{enumerate}
			\item $\OPT(J)=M$. 
			\item $\OPT(J)>M+\alpha\ls M$. 
		\end{enumerate}
	\end{theorem}
 Thus, unless $\mathrm P=\mathrm{NP}$,
	no deterministic polynomial time algorithm has additive error
	$o(\ls\OPT)$ on this instance class.
More broadly, we prove a universal polynomial-time reduction
preserving the additive configuration LP gap of arbitrary
instances up to a constant factor and an additive constant. Simultaneously and independently,~\cite{cryptoeprint:2026/2644} obtained the same asymptotic additive hardness but without restricting to items arbitrarily close to $1/3$.  

	\begin{theorem}\label{thm:universal}
		For every $\varepsilon>0$ there is a polynomial-time
		algorithm which, given any bin packing instance $I$, produces a bin packing instance $J$ of $3M$ items, where $M\ge1$ is an integer,
		with sizes in
		$(1/3-\varepsilon,1/3+\varepsilon)\cap(1/4,1/2)$
		such that
		$
		\LP(J)=M=O((\LP(I)+1)^4)$ and
		$$\OPT(I)-\LP(I)\le3+96(\OPT(J)-\LP(J)).
		$$
	\end{theorem}

	Eisenbrand et al.~\cite{EPR13} suggested that much of
the hardness of bin packing already appears in the
three-item setting (3-partition). Ayyadevara et al.~\cite{Ayy22}
explicitly attribute to Rothvoss the conjecture that
these are the hardest bin packing instances; see also
the related conjecture in~\cite{HR17}.
Theorem~\ref{thm:universal} establishes this
worst-case behavior for the additive integrality gap
of the configuration LP, up to constant factors and
an additive constant, even when all item sizes are
arbitrarily close to $1/3$.
In particular, we establish an unconditional additive
configuration-LP gap of $\Omega(\ls\OPT)$ for 3-partition instances, with sizes arbitrarily close to $1/3$. 

	\begin{theorem}\label{thm:gap}
		There is an absolute constant $\beta>0$ such that, for every fixed $\varepsilon>0$, there are arbitrarily large integers $M$
		and instances $J$ of $3M$ items in $
		(1/3-\varepsilon,1/3+\varepsilon)\cap(1/4,1/2)$ satisfying $\OPT(J)-\LP(J)\ge\beta\ls \OPT(J)$
	\end{theorem}

	Our integrality gap lower bound holds immediately for the more general
classical Gilmore--Gomory configuration LP~\cite{GG61}.

	%
	
	\paragraph{Proof overview.}
	
Our reduction allows additive lower bounds for bin packing to be
reduced to instances with all sizes arbitrarily close to $1/3$,
so each bin contains at most three items. The main difficulty is
to control arbitrary packings of the reduced instances, not only packings produced by
the construction. We must also preserve fractional packings to obtain
configuration LP gaps.
	
	Our reduction can be described in high level as follows. 
	We map an input instance $I$ (the {\em source}) to an output instance $J$
	(the {\em target}). 
	Source and target bins pack items of $I$ and $J$,
	respectively. 
	We encode each source bin by a path, with one step per item.
	The items may be ordered arbitrarily within a feasible bin, and the chosen
	order determines the {\em states} needed by the path.
	For an integer bin target $b\ge1$, choose a finite
	set of states $V$ with cardinality $|V|\ge|I|+b$ and a function $h:V\to[0,1]$.
	A state $v\in V$ is a distinct label with value $h(v)$, an upper bound
	on the total size of the items inserted to the bin so far.
	To represent each step of the path by a three-item bin, we use the following item classes.
	
	 Each state has two complementary items: one represents the load (an upper bound on the total size of items in a source bin) before inserting a source item, and the other represents the load after inserting a source item. A source item $i\in I$ is
	represented by a third item, so the triple associated with a transition of states
	$v\to w$, where $v,w\in V$, is feasible exactly when $h(v)+s_I(i)\le h(w)$.
	Two pairs of {\em endpoint} items start and end each path, while {\em filler} items
	cover unused states. With $M=|V|+b$, a partition into $b$ disjoint state
	paths therefore gives a packing of the $3M$ target items into $M$ bins.
	To decode an arbitrary target packing, we assign integer scores
that distinguish the intended triples from all others.
Since the scores sum to zero, a packing using $M+t$ bins
has at most $96t$ items outside these triples.
The remaining triples connect source items into paths,
each representing a feasible source bin.
Packing the discarded source items separately yields
at most $b+96t$ source bins.
	
	In particular, this reduction can be used to show two things. First, the hardness result of~\cite{OAI26}, which uses five items per bin, can be converted to three items per bin. Second, to obtain
	Theorem~\ref{thm:universal}, we compute a near-optimal fractional packing
	and restrict it to the large items. Its partial
	loads define the states; duplicating states distributes the fractional
	weights so that no state has weight above one. The removed items can
	then be reinserted with only a constant additive loss.
	
	Finally, we show how the size of the construction of~\cite{OAI26}
	grows with its additive gap. We optimize the parameters and allow the gap grow with the input
	while keeping the reduction polynomial-time. Combining this bound
	with our three-item reduction gives both the additive hardness and
	configuration-LP gap of order $\Omega(\ls\OPT)$.

	\paragraph{Organization.}
	Section~\ref{sec:reduction} gives the reduction to three items per bin and its two applications. Section~\ref{sec:source} bounds the size of the construction of~\cite{OAI26}. Section~\ref{sec:applications} proves the additive lower bounds.
	
	\section{Reduction to three-item bins}\label{sec:reduction}
	
	In this section, we give the main reduction to three items per bin.\footnote{
		The results and proofs in Sections~\ref{sec:reduction}--\ref{sec:applications}
		were developed through iterative interactions with ChatGPT; see the
		AI methodology statement at the end of the paper.} 
	Intuitively, in our reduction, we consider the process of inserting items into one bin. A {\em state} specifies an
	upper bound on the total size inserted so far. Moving to the next
	state inserts one item without exceeding the next bound.
	The reduction represents this step by three target items packed together.
	Different choices of states give Lemma~\ref{lem:reduction}
	and Theorem~\ref{thm:universal}.
	Throughout the technical sections, item sizes are rational and
	$\varepsilon>0$ is rational. For fixed real $\varepsilon>0$, choose
	a smaller positive rational.
	
	\subsection{Finite state reduction}
In this section, we give a reduction to three-item
	bin packing using a finite set of states to track
	the partial loads of source bins.
	For nonnegative weights $(w_U)_{U\in\mathcal F}$ on a finite family of sets,
	the \emph{coverage} of an element $u$ is $\sum_{U\in\mathcal F:\,u\in U}w_U$.
	The reduction takes an arbitrary bin packing instance $I$, an integer
	$b\ge1$, a finite set $V$ of \emph{states}, and a function
	$h:V\to[0,1]\cap\Q$. Let $n=|I|$. Fix these inputs for the
	definitions below; the two applications of the reduction will choose $V$ and $h$
	differently. We proceed with further definitions that will be used extensively throughout this section. 
	A \emph{state path} is a sequence
	$
	\pi=(v_0,i_1,v_1,\ldots,i_k,v_k),
	$
	where $k$ is an integer with $0\le k\le n$, $v_0,\ldots,v_k\in V$ are pairwise distinct,
	$i_1,\ldots,i_k\in I$ are pairwise distinct, and
	\begin{equation}\label{eq:transitions}
		h(v_j)\ge h(v_{j-1})+s_I(i_j)\qquad(1\le j\le k).
	\end{equation}
	A state path describes items that fit together in one source bin.
	We will encode it by three-item target bins and recover source bins
	by finding such paths.
	Write $V(\pi)=\{v_0,\ldots,v_k\}$ and
	$I(\pi)=\{i_1,\ldots,i_k\}$. By~\eqref{eq:transitions} we have
	$
	s_I(I(\pi))\le h(v_k)-h(v_0)\le1.
	$
	Two state paths are \emph{disjoint} if they share no state or item.
	Let $\mathcal P(I,V,h)$ be the set of these state paths, abbreviated
	$\mathcal P$ when $I,V,h$ are fixed. Nonnegative weights
	$x=(x_\pi)_{\pi\in\mathcal P}$ are \emph{feasible for $(I,b,V,h)$} if
	\begin{equation}\label{eq:state-fractional}
		\sum_{\pi\in\mathcal P}x_\pi=b,\qquad
		\sum_{\substack{\pi\in\mathcal P\\i\in I(\pi)}}x_\pi=1\quad(i\in I),\qquad
		\sum_{\substack{\pi\in\mathcal P\\v\in V(\pi)}}x_\pi\le1\quad(v\in V).
	\end{equation}
	
	The total path weight is $b$, each item belongs to paths of total
	weight one, and each state belongs to paths of total weight at most one.
	
	\Needspace{5\baselineskip}
	The next lemma turns state paths into three-item bins and obtains the following property: $t$ extra target bins require at most
	$O(t)$ extra bins in the original instance. This will be used for both the computational hardness result and the LP integrality gap result.

	\begin{lemma}\label{lem:states}
		There is a polynomial-time algorithm that, given a bin packing instance
		$I$, an integer $b\ge1$, a finite set $V$, a function
		$h:V\to[0,1]\cap\Q$, and a rational $\varepsilon>0$ with
		$|V|\ge|I|+b$, outputs a bin packing instance $J$ of $3M$ items,
		where $M=|V|+b$, with sizes in
		$(1/3-\varepsilon,1/3+\varepsilon)\cap(1/4,1/2)$, such that:
		\begin{enumerate}
			\item If there exists a feasible weight family for $(I,b,V,h)$,
			then $\LP(J)=M$.
			\item If there are pairwise disjoint paths
			$\pi_1,\ldots,\pi_b\in\mathcal P(I,V,h)$ such that
			$I(\pi_1),\ldots,I(\pi_b)$ is a partition of $I$, then $\OPT(J)=M$.
			\item Given a packing of $J$ into $M+t$ bins, for an integer $t\ge0$,
			one can construct in polynomial time a packing of $I$ into at
			most $b+96t$ bins.
		\end{enumerate}
	\end{lemma}
	
	\begin{proof}
		Set $M=|V|+b$ and $f=|V|-n-b$. All the following symbols denote
		distinct new items outside $I$; their classes are pairwise disjoint:
		\[
		\begin{aligned}
			\mathcal S^\delta&=\{S_v^\delta:v\in V\}
			&&(\delta\in\{-,+\}),\\
			\mathcal X&=\{X_i:i\in I\},\\
			\mathcal X^0&=\{X_r^0:r\in\Z,\ 1\le r\le f\},\\
			\mathcal E_\ell^\delta&=\{E_{j,\ell}^\delta:j\in\{1,\ldots,b\}\}
			&&(\delta\in\{-,+\},\ \ell\in\{1,2\}).
		\end{aligned}
		\]
		Let $J=\mathcal S^-\cup\mathcal S^+\cup\mathcal X\cup\mathcal X^0
		\cup\bigcup_{\delta\in\{-,+\},\,\ell\in\{1,2\}}\mathcal E_\ell^\delta$.
		Define $p=(p_1,p_2):J\to\Z\times\Q$. The state, source, and filler
		items have scores
		\begin{equation}\label{eq:state-scores}
			\begin{aligned}
				p(S_v^+)&=(-4,h(v))&&(v\in V),\\
				p(S_v^-)&=(-1,-h(v))&&(v\in V),\\
				p(X_i)&=(5,s_I(i))&&(i\in I),\\
				p(g)&=(5,0)&&(g\in\mathcal X^0).
			\end{aligned}
		\end{equation}
		The endpoint items have scores
		\begin{equation}\label{eq:endpoint-scores}
			\begin{aligned}
				p(E_{j,1}^-)&=(-3,0),\\
				p(E_{j,2}^-)&=(4,0),\\
				p(E_{j,1}^+)&=(1,0),\\
				p(E_{j,2}^+)&=(3,-1)
			\end{aligned}
			\qquad(1\le j\le b).
		\end{equation}
		Call $p_1$ the \emph{type score} and $p_2$ the \emph{load score};
		write $p_j(U)=\sum_{u\in U}p_j(u)$ for $U\subseteq J$ and $j\in\{1,2\}$.
		Set $\lambda=\min\{\varepsilon/6,1/100\}$ and
		\begin{equation}\label{eq:state-sizes}
			s_J(u)=\frac13+\lambda\left(p_1(u)+\frac{p_2(u)}4\right)
			\quad(u\in J).
		\end{equation}
		The \emph{class pattern} of $U\subseteq J$ is the multiset containing
		$|U\cap\mathcal A|$ copies of each of the eight classes $\mathcal A$;
		juxtaposition lists its elements. A \emph{filler triple} is
		$\{S_v^+,S_v^-,g\}$ with $v\in V$ and $g\in\mathcal X^0$.
		
		\begin{claim}\label{cl:state-size}
			$|J|=3M$, $p_1(J)=0$, and every size lies in
			$(1/3-\varepsilon,1/3+\varepsilon)\cap(1/4,1/2)$.
		\end{claim}
		\begin{claimproof}
			The item counts and scores give
			\begin{equation}\label{eq:state-count}
				|J|=2|V|+n+f+4b=3(|V|+b)=3M
			\end{equation}
			and
			\begin{equation}\label{eq:state-balance}
				p_1(J)=-5|V|+5(n+f)+5b=0.
			\end{equation}
			By~\eqref{eq:state-scores}, \eqref{eq:endpoint-scores}, and~\eqref{eq:state-sizes},
			$|s_J(u)-1/3|\le21\lambda/4<\min\{\varepsilon,1/12\}$ for every $u\in J$.
		\end{claimproof}
		
		\begin{claim}\label{cl:state-triples}
			Every feasible triple $U$ has $p_1(U)\le0$. The triples with $p_1(U)=0$
			have precisely the patterns in~\eqref{eq:state-patterns}; they are feasible iff $p_2(U)\le0$.
		\end{claim}
		\begin{claimproof}
			The only three-term zero sums, allowing repetition, from
			$\{-4,-3,-1,1,3,4,5\}$ are $-4-1+5$, $-3-1+4$, and $-4+1+3$.
			With two negative terms, their magnitudes must sum to $4$ or $5$
			(only $1+3$ or $1+4$); with one negative term, the positive
			terms must sum to $4$ (only $1+3$). Thus the patterns are
			\begin{equation}\label{eq:state-patterns}
				\begin{aligned}
					&\mathcal S^+\mathcal S^-\mathcal X,
					&&\mathcal S^+\mathcal S^-\mathcal X^0,\\
					&\mathcal S^-\mathcal E_1^-\mathcal E_2^-,
					&&\mathcal S^+\mathcal E_1^+\mathcal E_2^+.
				\end{aligned}
			\end{equation}
			For every triple $U$, we have $|p_2(U)|\le3$ and
			$s_J(U)=1+\lambda(p_1(U)+p_2(U)/4)$.
			The assertions follow from $p_1(U)\in\Z$.
		\end{claimproof}
		
		Figure~\ref{fig:transition} shows all target bins for a source bin
		with two items and one unused state.
		
		\begin{figure}[htbp]
			\centering
			\begin{tikzpicture}[font=\small]
				\node at (-1.9,.775) {$I$};
				\fill[bpx] (-1.42000000,0.450) rectangle (-0.71000000,1.100);
				\node at (-1.06500000,0.775) {$1$};
				\node at (-1.06500000,0.170) {$\frac{1}{4}$};
				\fill[bpx] (-0.71000000,0.450) rectangle (0.23666667,1.100);
				\node at (-0.23666667,0.775) {$2$};
				\node at (-0.23666667,0.170) {$\frac{1}{3}$};
				\draw[bpedge] (-0.71000000,0.450)--(-0.71000000,1.100);
				\draw[bpedge] (0.23666667,0.450)--(0.23666667,1.100);
				\draw[bpedge] (-1.42000000,0.450) rectangle (1.42000000,1.100);
				\draw[bpedge] (-1.42000000,1.10)--(-1.42000000,1.18);
				\node[above,inner sep=2pt] at (-1.42000000,1.18) {$0$};
				\draw[bpedge] (-0.71000000,1.10)--(-0.71000000,1.18);
				\node[above,inner sep=2pt] at (-0.71000000,1.18) {$\frac{1}{4}$};
				\draw[bpedge] (0.23666667,1.10)--(0.23666667,1.18);
				\node[above,inner sep=2pt] at (0.23666667,1.18) {$\frac{7}{12}$};
				\draw[bpedge] (1.42000000,1.10)--(1.42000000,1.18);
				\node[above,inner sep=2pt] at (1.42000000,1.18) {$1$};
				\node at (-7.85,-.725) {$J$};
				\fill[bpsm] (-7.42000000,-1.050) rectangle (-6.50173333,-0.400);
				\node at (-6.96086667,-0.725) {$S_0^-$};
				\node at (-6.96086667,-1.330) {$\frac{1552}{4800}$};
				\fill[bpemone] (-6.50173333,-1.050) rectangle (-5.64026667,-0.400);
				\node at (-6.07100000,-0.725) {$E_{1,1}^-$};
				\node at (-6.07100000,-1.330) {$\frac{1456}{4800}$};
				\draw[bpedge] (-6.50173333,-1.050)--(-6.50173333,-0.400);
				\fill[bpemtwo] (-5.64026667,-1.050) rectangle (-4.58000000,-0.400);
				\node at (-5.11013333,-0.725) {$E_{1,2}^-$};
				\node at (-5.11013333,-1.330) {$\frac{1792}{4800}$};
				\draw[bpedge] (-5.64026667,-1.050)--(-5.64026667,-0.400);
				\draw[bpedge] (-7.42000000,-1.050) rectangle (-4.58000000,-0.400);
				\fill[bpsp] (-4.42000000,-1.050) rectangle (-3.58693333,-0.400);
				\node at (-4.00346667,-0.725) {$S_0^+$};
				\node at (-4.00346667,-1.330) {$\frac{1408}{4800}$};
				\fill[bpx] (-3.58693333,-1.050) rectangle (-2.49649167,-0.400);
				\node at (-3.04171250,-0.725) {$X_1$};
				\node at (-3.04171250,-1.330) {$\frac{1843}{4800}$};
				\draw[bpedge] (-3.58693333,-1.050)--(-3.58693333,-0.400);
				\fill[bpsm] (-2.49649167,-1.050) rectangle (-1.58000000,-0.400);
				\node at (-2.03824583,-0.725) {$S_1^-$};
				\node at (-2.03824583,-1.330) {$\frac{1549}{4800}$};
				\draw[bpedge] (-2.49649167,-1.050)--(-2.49649167,-0.400);
				\draw[bpedge] (-4.42000000,-1.050) rectangle (-1.58000000,-0.400);
				\fill[bpsp] (-1.42000000,-1.050) rectangle (-0.58515833,-0.400);
				\node at (-1.00257917,-0.725) {$S_1^+$};
				\node at (-1.00257917,-1.330) {$\frac{1411}{4800}$};
				\fill[bpx] (-0.58515833,-1.050) rectangle (0.50587500,-0.400);
				\node at (-0.03964167,-0.725) {$X_2$};
				\node at (-0.03964167,-1.330) {$\frac{1844}{4800}$};
				\draw[bpedge] (-0.58515833,-1.050)--(-0.58515833,-0.400);
				\fill[bpsm] (0.50587500,-1.050) rectangle (1.42000000,-0.400);
				\node at (0.96293750,-0.725) {$S_2^-$};
				\node at (0.96293750,-1.330) {$\frac{1545}{4800}$};
				\draw[bpedge] (0.50587500,-1.050)--(0.50587500,-0.400);
				\draw[bpedge] (-1.42000000,-1.050) rectangle (1.42000000,-0.400);
				\fill[bpsp] (1.58000000,-1.050) rectangle (2.41720833,-0.400);
				\node at (1.99860417,-0.725) {$S_2^+$};
				\node at (1.99860417,-1.330) {$\frac{1415}{4800}$};
				\fill[bpepone] (2.41720833,-1.050) rectangle (3.39227500,-0.400);
				\node at (2.90474167,-0.725) {$E_{1,1}^+$};
				\node at (2.90474167,-1.330) {$\frac{1648}{4800}$};
				\draw[bpedge] (2.41720833,-1.050)--(2.41720833,-0.400);
				\fill[bpeptwo] (3.39227500,-1.050) rectangle (4.41704167,-0.400);
				\node at (3.90465833,-0.725) {$E_{1,2}^+$};
				\node at (3.90465833,-1.330) {$\frac{1732}{4800}$};
				\draw[bpedge] (3.39227500,-1.050)--(3.39227500,-0.400);
				\draw[bpedge] (4.41704167,-1.050)--(4.41704167,-0.400);
				\draw[bpedge] (1.58000000,-1.050) rectangle (4.42000000,-0.400);
				\fill[bpsp] (4.58000000,-1.050) rectangle (5.41839167,-0.400);
				\node at (4.99919583,-0.725) {$S_3^+$};
				\node at (4.99919583,-1.330) {$\frac{1417}{4800}$};
				\fill[bpsm] (5.41839167,-1.050) rectangle (6.33133333,-0.400);
				\node at (5.87486250,-0.725) {$S_3^-$};
				\node at (5.87486250,-1.330) {$\frac{1543}{4800}$};
				\draw[bpedge] (5.41839167,-1.050)--(5.41839167,-0.400);
				\fill[bpf] (6.33133333,-1.050) rectangle (7.42000000,-0.400);
				\node at (6.87566667,-0.725) {$X_1^0$};
				\node at (6.87566667,-1.330) {$\frac{1840}{4800}$};
				\draw[bpedge] (6.33133333,-1.050)--(6.33133333,-0.400);
				\draw[bpedge] (4.58000000,-1.050) rectangle (7.42000000,-0.400);
			\end{tikzpicture}
			\caption{A complete packing for $I=\{1,2\}$ with sizes $1/4,1/3$.
				Use $b=1$, $V=\{0,1,2,3\}$, and
				$(h(0),h(1),h(2),h(3))=(0,1/4,7/12,3/4)$.
				The top bin shows the source items and partial loads. The five bottom
				bins contain the packing of the corresponding target instance. 
				Here $\varepsilon=1/10$, so $\lambda=1/100$.}
			\label{fig:transition}
		\end{figure}
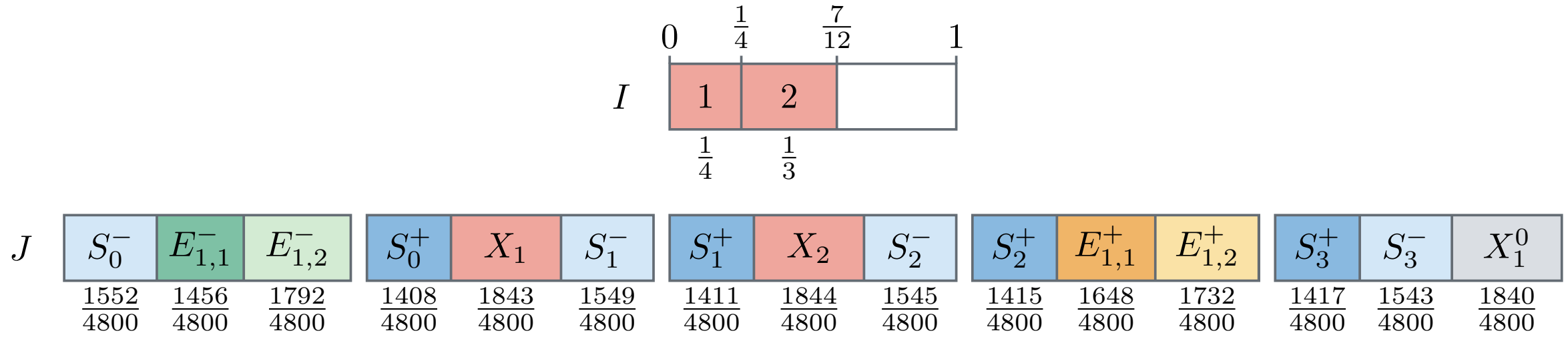
		
		For $\pi=(v_0,i_1,v_1,\ldots,i_k,v_k)\in\mathcal P$ and $1\le j\le b$,
		define
		\begin{equation}\label{eq:path-lift}
			\mathcal T_j(\pi)=
			\{\{S_{v_0}^-,E_{j,1}^-,E_{j,2}^-\},\{S_{v_k}^+,E_{j,1}^+,E_{j,2}^+\}\}
			\cup\{\{S_{v_{\ell-1}}^+,X_{i_\ell},S_{v_\ell}^-\}:1\le\ell\le k\}.
		\end{equation}
		\begin{claim}\label{cl:state-integral}
			Every $\mathcal T_j(\pi)$ consists of feasible triples. Disjoint paths
			$\pi_1,\ldots,\pi_b$ whose item sets partition $I$ imply $\OPT(J)=M$.
		\end{claim}
		\begin{claimproof}
			The triples in~\eqref{eq:path-lift} have type score zero and load scores
			$-h(v_0)$, $h(v_k)-1$, and $h(v_{\ell-1})+s_I(i_\ell)-h(v_\ell)$.
			These are nonpositive by~\eqref{eq:transitions} and $h(V)\subseteq[0,1]$.
			For the disjoint paths, let $V_0=V\setminus\bigcup_{j=1}^b V(\pi_j)$.
			Since $\sum_{j=1}^b|V(\pi_j)|=n+b$, we have $|V_0|=f$.
			Choose a bijection $g:V_0\to\mathcal X^0$. The family
			\[
			\mathcal B_J=\bigcup_{j=1}^b\mathcal T_j(\pi_j)
			\cup\{\{S_v^+,S_v^-,g(v)\}:v\in V_0\}
			\]
			partitions $J$ into feasible triples: each added filler triple has both
			scores zero. Hence $|\mathcal B_J|=M$. No bin holds four items by
			Claim~\ref{cl:state-size}, so $\OPT(J)=M$.
		\end{claimproof}
		
		\begin{claim}\label{cl:state-fractional}
			A feasible weight family for $(I,b,V,h)$ implies $\LP(J)=M$.
		\end{claim}
		\begin{claimproof}
			Fix such a family $x$. For every $U\in\mathcal H(J)$, initialize
			\[
			y_U=\frac1b\sum_{\pi\in\mathcal P}
			\sum_{\substack{1\le j\le b\\U\in\mathcal T_j(\pi)}}x_\pi.
			\]
			Equation~\eqref{eq:state-fractional} gives coverage one for every
			$X_i$ and endpoint item. Each $S_v^\pm$ has coverage
			$\sum_{\pi:\,v\in V(\pi)}x_\pi\le1$. Since $|V(\pi)|=|I(\pi)|+1$,
			\[
			\sum_{v\in V}\sum_{\pi:\,v\in V(\pi)}x_\pi
			=\sum_{\pi\in\mathcal P}x_\pi(|I(\pi)|+1)=n+b.
			\]
			Thus $\sum_{v\in V}(1-\sum_{\pi:\,v\in V(\pi)}x_\pi)=f$.
			If $f>0$, perform the assignments
			\[
			y_{\{S_v^+,S_v^-,g\}}\leftarrow y_{\{S_v^+,S_v^-,g\}}
			+\frac{1-\sum_{\pi:\,v\in V(\pi)}x_\pi}{f}
			\quad(v\in V,\ g\in\mathcal X^0).
			\]
			Each state item's coverage becomes one, and each filler has coverage
			$\frac1f\sum_{v\in V}(1-\sum_{\pi:\,v\in V(\pi)}x_\pi)=1$.
			If $f=0$, the nonnegative deficits sum to zero, so every state item
			already has coverage one and $\mathcal X^0=\varnothing$.
			All positive-weight configurations are feasible triples by
			Claim~\ref{cl:state-integral} and the zero scores of filler triples. Hence
			\[
			3\sum_{U\in\mathcal H(J)}y_U
			=\sum_{u\in J}\sum_{U\ni u}y_U=|J|=3M.
			\]
			Conversely, $|U|\le3$ for every configuration $U$, so summing the
			LP constraints gives the inequality $3\LP(J)\ge|J|=3M$.
		\end{claimproof}
		
		\begin{claim}\label{cl:state-exceptions}
			For $t\in\Z_{\ge0}$, every packing $\mathcal B$ of $J$ into $M+t$ bins has a computable
			subfamily $\mathcal B_*$ with $|J\setminus\bigcup\mathcal B_*|\le96t$
			and $|U|=3$, $p_1(U)=0$ for every $U\in\mathcal B_*$.
		\end{claim}
		\begin{claimproof}
			Set $J_0=\bigcup_{U\in\mathcal B:\,|U|<3}U$,
			$\mathcal B_3=\{U\in\mathcal B:|U|=3\}$, and
			$$\mathcal B_-=\{U\in\mathcal B_3:p_1(U)<0\}.$$
			By~\eqref{eq:state-count},
			$
			|J_0|\le2\sum_{U\in\mathcal B}(3-|U|)=6t.
			$
			Claim~\ref{cl:state-triples}, integrality of $p_1$, and~\eqref{eq:state-balance} give
			\[
			|\mathcal B_-|\le-\sum_{U\in\mathcal B_3}p_1(U)
			=p_1(J_0)\le5|J_0|.
			\]
			Set $\mathcal B_* =\mathcal B_3\setminus\mathcal B_-$ and
			$J_*=J\setminus\bigcup\mathcal B_*=J_0\cup\bigcup_{U\in\mathcal B_-}U$.
			Then $|U|=3$ and $p_1(U)=0$ for all $U\in\mathcal B_*$, and
			$
			|J_*|=|J_0|+3|\mathcal B_-|\le16|J_0|\le96t.
			$
		\end{claimproof}
		
		\begin{claim}\label{cl:state-decode}
			For $t\in\Z_{\ge0}$, a packing of $J$ into $M+t$ bins yields, in polynomial time,
			a packing of $I$ into at most $b+96t$ bins.
		\end{claim}
		\begin{claimproof}
			Take $\mathcal B_*$ from Claim~\ref{cl:state-exceptions} and set
			$J_*=J\setminus\bigcup\mathcal B_*$.
			Define the directed graph $H=(V,E_H)$, allowing loops, by
			\[
			E_H=\{(v,w)\in V^2:\ \exists z\in\mathcal X\cup\mathcal X^0,
			\ \{S_v^+,z,S_w^-\}\in\mathcal B_*\}.
			\]
			For $(v,w)\in E_H$, its unique middle item $z$ defines the label
			\[
			\ell_H(v,w)=
			\begin{cases}
				i,&z=X_i\text{ for }i\in I,\\
				z,&z\in\mathcal X^0.
			\end{cases}
			\]
			Each $S_v^+$ and $S_v^-$ occurs in at most one member of $\mathcal B_*$,
			so every vertex has outdegree and indegree at most one.
			By~\eqref{eq:state-scores} and Claim~\ref{cl:state-triples},
			\begin{equation}\label{eq:edge-load}
				h(w)-h(v)\ge
				\begin{cases}
					s_I(\ell_H(v,w)),&\ell_H(v,w)\in I,\\
					0,&\ell_H(v,w)\in\mathcal X^0
				\end{cases}
				\quad((v,w)\in E_H).
			\end{equation}
			Every weakly connected component is an isolated vertex, a directed
			path, or a directed cycle. For a path or cycle $\gamma$, let
			$I(\gamma)=\{\ell_H(e):e\in E(\gamma)\}\cap I$, where $E(\gamma)$ is its edge set.
			Summing~\eqref{eq:edge-load} gives
			\[
			s_I(I(\gamma))\le
			\begin{cases}
				0,&\gamma\text{ is a cycle},\\
				h(w)-h(v)\le1,&\gamma\text{ is a path from }v\text{ to }w.
			\end{cases}
			\]
			Positive source sizes imply $I(\gamma)=\varnothing$ for cycles.
			Let $\Gamma$ be the directed path components with at least one edge.
			Then
			\[
			\mathcal B_I=
			\{I(\gamma):\gamma\in\Gamma,\ I(\gamma)\ne\varnothing\}
			\cup\{\{i\}:i\in I,\ X_i\in J_*\}
			\]
			partitions $I$ into sets of total size at most one.
			The initial vertex $v$ of each $\gamma\in\Gamma$ has indegree zero.
			By~\eqref{eq:state-patterns}, either $S_v^-\in J_*$ or its bin in
			$\mathcal B_*$ has pattern $\mathcal S^-\mathcal E_1^-\mathcal E_2^-$;
			otherwise that bin gives an incoming edge. There are at most $b$
			bins of this pattern, and distinct paths have distinct initial vertices.
			Consequently,
			\[
			|\mathcal B_I|\le b+|J_*\cap\mathcal S^-|+|J_*\cap\mathcal X|
			\le b+|J_*|\le b+96t.
			\]
			Computing $\mathcal B_*$, $H$, and its components takes polynomial time.
		\end{claimproof}
		Claims~\ref{cl:state-size}, \ref{cl:state-integral},
		\ref{cl:state-fractional}, and~\ref{cl:state-decode} prove the lemma.
		The construction uses $O(|V|+b)$ items with polynomial binary length.
	\end{proof}

	We next specialize Lemma~\ref{lem:states} to instances with five
	equally sized item classes.
	\subsection{Five item classes}
Now, we show that the reduction preserves integral and fractional partitions into feasible bins each with five items.
	For this subsection, fix an integer $N\ge1$ and an instance $I$
	with a specified partition
	$I=A\mathbin{\dot\cup}B\mathbin{\dot\cup}C
	\mathbin{\dot\cup}D\mathbin{\dot\cup}E$ satisfying
	$|A|=|B|=|C|=|D|=|E|=N$. These five disjoint subsets are the
	item classes. A \emph{feasible five-tuple} is
	$(a,b,c,d,e)\in A\times B\times C\times D\times E$ satisfying
	$s_I(a)+s_I(b)+s_I(c)+s_I(d)+s_I(e)\le1$.
	A partition into feasible five-tuples means that their underlying
	five-element sets partition $I$. A \emph{fractional partition}
	assigns nonnegative weights to feasible five-tuples, so that each
	item belongs to tuples of total weight one.
	\Needspace{5\baselineskip}
	The next lemma retains all feasible prefixes of at most four items,
	so that integral and fractional five-tuple partitions give state paths.
	
	\begin{lemma}\label{lem:reduction}
		Let $N\ge1$ be an integer and let $I$ have five classes of $N$ items each.
		For every fixed $\varepsilon>0$, one can construct in polynomial
		time an instance $J$ of $3M$ items with sizes in
		$(1/3-\varepsilon,1/3+\varepsilon)\cap(1/4,1/2)$,
		where $M=O(N^4)$, such that:
		\begin{enumerate}
			\item A partition of $I$ into feasible five-tuples implies
			$\OPT(J)=M$.
			\item A fractional partition of $I$ into feasible five-tuples
			implies $\LP(J)=M$.
			\item For every integer $t\ge0$, a packing of $J$ into $M+t$ bins
			can be converted in polynomial time into a packing of $I$ into at most
			$N+C_0t$ bins, where $C_0=96$.
		\end{enumerate}
	\end{lemma}
	
	\begin{proof}
		Set $b=N$ and $(A_1,A_2,A_3,A_4,A_5)=(A,B,C,D,E)$. Define
		\begin{equation}\label{eq:five-prefixes}
			\mathcal F_j=\left\{(i_1,\ldots,i_j)\in\prod_{\ell=1}^j A_\ell:
			\sum_{\ell=1}^j s_I(i_\ell)\le1\right\}
			\quad(1\le j\le5).
		\end{equation}
		Use labels $u_r=(0,r)$ and $w_r=(1,r)$ for $1\le r\le N$,
		and $v(i_1,\ldots,i_j)=(2,j,(i_1,\ldots,i_j))$ for
		$(i_1,\ldots,i_j)\in\mathcal F_j$, $1\le j\le4$. Set
		\[
		V_0=\{u_r,w_r:1\le r\le N\}
		\cup\bigcup_{j=1}^4\{v(i_1,\ldots,i_j):(i_1,\ldots,i_j)\in\mathcal F_j\}.
		\]
		Let $q=\max\{0,6N-|V_0|\}$ and define $q$ padding states
		$z_r=(3,r)$, $1\le r\le q$. The state set is
		\begin{equation}\label{eq:five-states}
			V=V_0\mathbin{\dot\cup}\{z_r:1\le r\le q\}.
		\end{equation}
		Define the loads by
		\begin{equation}\label{eq:five-loads}
			\begin{aligned}
				h(u_r)&=0&&(1\le r\le N),\\
				h(w_r)&=1&&(1\le r\le N),\\
				h(z_r)&=0&&(1\le r\le q),\\
				h(v(i_1,\ldots,i_j))&=\sum_{\ell=1}^j s_I(i_\ell)
				&&(1\le j\le4,\ (i_1,\ldots,i_j)\in\mathcal F_j).
			\end{aligned}
		\end{equation}
		For $t=(i_1,\ldots,i_5)\in\mathcal F_5$ and $1\le r,s\le N$, define
		\begin{equation}\label{eq:five-paths}
			\begin{aligned}
				\pi_t^{rs}&=(v_0,i_1,v_1,\ldots,i_5,v_5),\\
				v_0&=u_r,\quad v_5=w_s,\quad v_j=v(i_1,\ldots,i_j)\quad(1\le j\le4).
			\end{aligned}
		\end{equation}
		Apply the construction of Lemma~\ref{lem:states} to $(I,N,V,h)$
		to obtain $J$, with $M=|V|+N$.
		
		\begin{claim}\label{cl:five-states}
			$6N\le|V|=O(N^4)$, $h(V)\subseteq[0,1]\cap\Q$, and
			$\pi_t^{rs}\in\mathcal P(I,V,h)$ for every $t\in\mathcal F_5$ and $1\le r,s\le N$.
		\end{claim}
		\begin{claimproof}
			Equations~\eqref{eq:five-prefixes}--\eqref{eq:five-states} give
			$|V|=\max\{6N,2N+\sum_{j=1}^4|\mathcal F_j|\}=O(N^4)$.
			Equation~\eqref{eq:five-loads} gives $h(V)\subseteq[0,1]\cap\Q$.
			The state tags and disjoint item classes make all states and items
			within~\eqref{eq:five-paths} distinct. The first four transition
			inequalities are equalities; the last is
			$h(v_4)+s_I(i_5)=\sum_{j=1}^5s_I(i_j)\le1=h(v_5)$.
		\end{claimproof}
		
		\begin{claim}\label{cl:five-integral}
			A partition into feasible five-tuples gives $N$ disjoint state paths
			whose item sets partition~$I$.
		\end{claim}
		\begin{claimproof}
			Enumerate the tuples as $t_1,\ldots,t_N$ and use $\pi_{t_j}^{jj}$,
			$1\le j\le N$, from~\eqref{eq:five-paths}.
			Different tuples have different first items, hence disjoint prefix states;
			their initial and final states are also distinct.
		\end{claimproof}
		
		\begin{claim}\label{cl:five-fractional}
			A fractional partition into feasible five-tuples gives a feasible
			weight family for $(I,N,V,h)$.
		\end{claim}
		\begin{claimproof}
			Let $x_t$ be the weight of $t\in\mathcal F_5$. Give each
			$\pi_t^{rs}$ weight $x_t/N^2$ and every other path weight zero.
			Since each tuple contains one item of $A$, we have
			$\sum_{t\in\mathcal F_5}x_t=N$. Thus the path weights sum to $N$,
			and each item belongs to paths of total weight one.
			Each $u_r,w_s$ belongs to paths of total weight
			$\sum_{t\in\mathcal F_5}x_t/N=1$.
			For a prefix state $v(i_1,\ldots,i_j)$, this weight is at most
			$\sum_{t\in\mathcal F_5:\,t_1=i_1}x_t=1$, where $t_1$ denotes the first
			entry of $t$; padding states belong to no positive-weight path.
			These are exactly the conditions in~\eqref{eq:state-fractional}.
		\end{claimproof}
		Claims~\ref{cl:five-states}--\ref{cl:five-fractional} and
		Lemma~\ref{lem:states} give (i)--(iii), with $C_0=96$.
		The state set has polynomial size and can be enumerated in polynomial time.
	\end{proof}
	
	\subsection{Universal configuration LP gap reduction}
	We show that every bin packing instance has a three-item corresponding instance preserving its configuration-LP gap up to constant factors.
	We use the large items and create states for the partial loads
	of a fractional packing.
	\begin{lemma}\label{lem:universal-decode}
		For the reduction in Theorem~\ref{thm:universal}, a packing of $J$ into
		$\LP(J)+t$ bins can be converted in polynomial time into a packing of
		$I$ into at most $\LP(I)+3+96t$ bins, for every integer $t\ge0$.
	\end{lemma}
	
	\begin{proof}[Proof of Theorem~\ref{thm:universal} and Lemma~\ref{lem:universal-decode}]
		If $I=\varnothing$, set $J=\{1,2,3\}$, $s_J(j)=1/3$ for $j\in J$,
		and $M=1$; the decoder returns the empty packing.
		Otherwise, set $n=|I|$ and $L=\LP(I)$.
		Apply the configuration-LP approximation algorithm of~\cite{Sha20}
		with accuracy $1/(2n)$ to obtain a rational feasible solution
		$(\widehat y_U)_{U\in\mathcal H(I)}$ with polynomially many positive coordinates and cost
		$W\le(1+1/(2n))L$. Set
		$
		B=\lceil W\rceil$,
		$R=\{i\in I:s_I(i)\ge1/(B+1)\}$, and $m=|R|$.
		Initialize weights on $\mathcal H(R)$ by
		\[
		y_C=\sum_{\substack{U\in\mathcal H(I)\\U\cap R=C}}\widehat y_U
		\quad(C\in\mathcal H(R)).
		\]
		Process each $i\in R$ once. Enumerate the configurations containing
		$i$ with positive weight as $C_1,\ldots,C_a$, set
		$e_i=\sum_{j=1}^a y_{C_j}-1$, and define
		\begin{equation}\label{eq:universal-trim}
			\delta_{C_j}=\min\left\{y_{C_j},\max\left\{0,e_i-\sum_{\ell=1}^{j-1}y_{C_\ell}\right\}\right\}
			\quad(1\le j\le a).
		\end{equation}
		Using the weights before the update on all right-hand sides, set simultaneously
		\begin{equation}\label{eq:universal-update}
			\begin{aligned}
				y_C&\leftarrow y_C-\delta_C,\\
				y_{C\setminus\{i\}}&\leftarrow y_{C\setminus\{i\}}+\delta_C
			\end{aligned}
			\qquad(C\in\{C_1,\ldots,C_a\}).
		\end{equation}
		Let $\mathcal D=\{C\ne\varnothing:y_C>0\}$ after these updates.
		Compute an optimal basic solution $x$ of the following LP
		\begin{equation}\label{eq:universal-exact-lp}
			\min\left\{\sum_{C\in\mathcal D}x_C:
			x_C\ge0\ (C\in\mathcal D),\quad
			\sum_{\substack{C\in\mathcal D\\i\in C}}x_C=1\ (i\in R)\right\}.
		\end{equation}
		For $m=0$, use the zero solution. Set
		$x_\varnothing=B-\sum_{C\in\mathcal D}x_C$ and
		$\mathcal C=\{C\in\mathcal D\cup\{\varnothing\}:x_C>0\}$.
		For each $C\in\mathcal C$, fix an ordering $(i_1,\ldots,i_k)$,
		where $k=|C|$, and define $h_0(C)=0$ and
		$h_j(C)=\sum_{\ell=1}^j s_I(i_\ell)$ for $1\le j\le k$.
		Set $H=\bigcup_{C\in\mathcal C}\{h_0(C),\ldots,h_{|C|}(C)\}$ and
		\[
		\mu_u=\sum_{\substack{C\in\mathcal C\\
				u\in\{h_0(C),\ldots,h_{|C|}(C)\}}}x_C\quad(u\in H).
		\]
		Thus $\mu_u$ is the total weight of configurations having partial load $u$.
		Define
		\begin{equation}\label{eq:universal-states}
			V=\{(u,r):u\in H,\ r\in\{1,\ldots,\lceil\mu_u\rceil\}\},
		\end{equation}
        and define $h(u,r)=u$ for $(u,r)\in V$. 
		Apply the construction of Lemma~\ref{lem:states} to $(R,B,V,h)$
		to obtain $J$, with $M=|V|+B$.
		Figure~\ref{fig:state-copies} shows state copies and the filler
		configurations that complete their fractional coverage.
		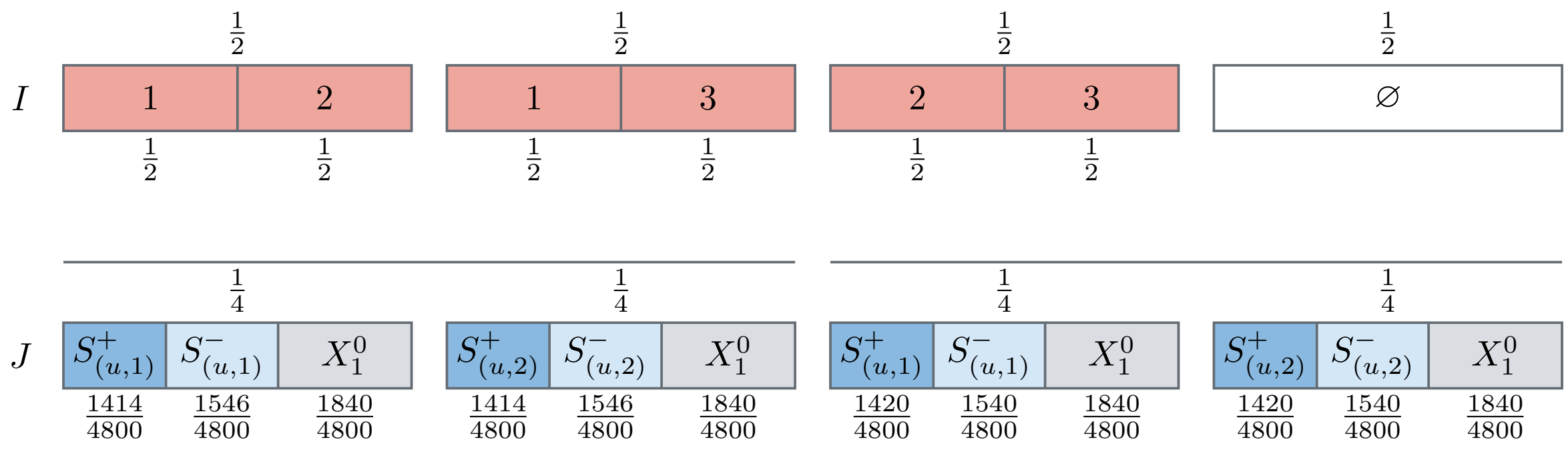
\begin{figure}[htbp]
			\centering
			\begin{tikzpicture}[font=\small]
				\node at (-7.85,.375) {$I$};
				\fill[bpx] (-7.42500000,0.050) rectangle (-5.70000000,0.700);
				\node at (-6.56250000,0.375) {$1$};
				\node at (-6.56250000,-0.230) {$\frac{1}{2}$};
				\fill[bpx] (-5.70000000,0.050) rectangle (-3.97500000,0.700);
				\node at (-4.83750000,0.375) {$2$};
				\node at (-4.83750000,-0.230) {$\frac{1}{2}$};
				\draw[bpedge] (-5.70000000,0.050)--(-5.70000000,0.700);
				\draw[bpedge] (-7.42500000,0.050) rectangle (-3.97500000,0.700);
				\node at (-5.700,1.020) {$\frac{1}{2}$};
				\fill[bpx] (-3.62500000,0.050) rectangle (-1.90000000,0.700);
				\node at (-2.76250000,0.375) {$1$};
				\node at (-2.76250000,-0.230) {$\frac{1}{2}$};
				\fill[bpx] (-1.90000000,0.050) rectangle (-0.17500000,0.700);
				\node at (-1.03750000,0.375) {$3$};
				\node at (-1.03750000,-0.230) {$\frac{1}{2}$};
				\draw[bpedge] (-1.90000000,0.050)--(-1.90000000,0.700);
				\draw[bpedge] (-3.62500000,0.050) rectangle (-0.17500000,0.700);
				\node at (-1.900,1.020) {$\frac{1}{2}$};
				\fill[bpx] (0.17500000,0.050) rectangle (1.90000000,0.700);
				\node at (1.03750000,0.375) {$2$};
				\node at (1.03750000,-0.230) {$\frac{1}{2}$};
				\fill[bpx] (1.90000000,0.050) rectangle (3.62500000,0.700);
				\node at (2.76250000,0.375) {$3$};
				\node at (2.76250000,-0.230) {$\frac{1}{2}$};
				\draw[bpedge] (1.90000000,0.050)--(1.90000000,0.700);
				\draw[bpedge] (0.17500000,0.050) rectangle (3.62500000,0.700);
				\node at (1.900,1.020) {$\frac{1}{2}$};
				\node at (5.700,0.375) {$\varnothing$};
				\draw[bpedge] (3.97500000,0.050) rectangle (7.42500000,0.700);
				\node at (5.700,1.020) {$\frac{1}{2}$};
				\draw[bpedge] (-7.425,-1.25)--(-.175,-1.25);
				\draw[bpedge] (.175,-1.25)--(7.425,-1.25);
				\node at (-7.85,-2.175) {$J$};
				\fill[bpsp] (-7.42500000,-2.500) rectangle (-6.40868750,-1.850);
				\node at (-6.91684375,-2.175) {$S_{(u,1)}^+$};
				\node at (-6.91684375,-2.780) {$\frac{1414}{4800}$};
				\fill[bpsm] (-6.40868750,-2.500) rectangle (-5.29750000,-1.850);
				\node at (-5.85309375,-2.175) {$S_{(u,1)}^-$};
				\node at (-5.85309375,-2.780) {$\frac{1546}{4800}$};
				\draw[bpedge] (-6.40868750,-2.500)--(-6.40868750,-1.850);
				\fill[bpf] (-5.29750000,-2.500) rectangle (-3.97500000,-1.850);
				\node at (-4.63625000,-2.175) {$X_1^0$};
				\node at (-4.63625000,-2.780) {$\frac{1840}{4800}$};
				\draw[bpedge] (-5.29750000,-2.500)--(-5.29750000,-1.850);
				\draw[bpedge] (-7.42500000,-2.500) rectangle (-3.97500000,-1.850);
				\node at (-5.700,-1.530) {$\frac{1}{4}$};
				\fill[bpsp] (-3.62500000,-2.500) rectangle (-2.60868750,-1.850);
				\node at (-3.11684375,-2.175) {$S_{(u,2)}^+$};
				\node at (-3.11684375,-2.780) {$\frac{1414}{4800}$};
				\fill[bpsm] (-2.60868750,-2.500) rectangle (-1.49750000,-1.850);
				\node at (-2.05309375,-2.175) {$S_{(u,2)}^-$};
				\node at (-2.05309375,-2.780) {$\frac{1546}{4800}$};
				\draw[bpedge] (-2.60868750,-2.500)--(-2.60868750,-1.850);
				\fill[bpf] (-1.49750000,-2.500) rectangle (-0.17500000,-1.850);
				\node at (-0.83625000,-2.175) {$X_1^0$};
				\node at (-0.83625000,-2.780) {$\frac{1840}{4800}$};
				\draw[bpedge] (-1.49750000,-2.500)--(-1.49750000,-1.850);
				\draw[bpedge] (-3.62500000,-2.500) rectangle (-0.17500000,-1.850);
				\node at (-1.900,-1.530) {$\frac{1}{4}$};
				\fill[bpsp] (0.17500000,-2.500) rectangle (1.19562500,-1.850);
				\node at (0.68531250,-2.175) {$S_{(u,1)}^+$};
				\node at (0.68531250,-2.780) {$\frac{1420}{4800}$};
				\fill[bpsm] (1.19562500,-2.500) rectangle (2.30250000,-1.850);
				\node at (1.74906250,-2.175) {$S_{(u,1)}^-$};
				\node at (1.74906250,-2.780) {$\frac{1540}{4800}$};
				\draw[bpedge] (1.19562500,-2.500)--(1.19562500,-1.850);
				\fill[bpf] (2.30250000,-2.500) rectangle (3.62500000,-1.850);
				\node at (2.96375000,-2.175) {$X_1^0$};
				\node at (2.96375000,-2.780) {$\frac{1840}{4800}$};
				\draw[bpedge] (2.30250000,-2.500)--(2.30250000,-1.850);
				\draw[bpedge] (0.17500000,-2.500) rectangle (3.62500000,-1.850);
				\node at (1.900,-1.530) {$\frac{1}{4}$};
				\fill[bpsp] (3.97500000,-2.500) rectangle (4.99562500,-1.850);
				\node at (4.48531250,-2.175) {$S_{(u,2)}^+$};
				\node at (4.48531250,-2.780) {$\frac{1420}{4800}$};
				\fill[bpsm] (4.99562500,-2.500) rectangle (6.10250000,-1.850);
				\node at (5.54906250,-2.175) {$S_{(u,2)}^-$};
				\node at (5.54906250,-2.780) {$\frac{1540}{4800}$};
				\draw[bpedge] (4.99562500,-2.500)--(4.99562500,-1.850);
				\fill[bpf] (6.10250000,-2.500) rectangle (7.42500000,-1.850);
				\node at (6.76375000,-2.175) {$X_1^0$};
				\node at (6.76375000,-2.780) {$\frac{1840}{4800}$};
				\draw[bpedge] (6.10250000,-2.500)--(6.10250000,-1.850);
				\draw[bpedge] (3.97500000,-2.500) rectangle (7.42500000,-1.850);
				\node at (5.700,-1.530) {$\frac{1}{4}$};
			\end{tikzpicture}
			\caption{Fractional filler configurations for three source items
of size $1/2$. The four top configurations each have weight $1/2$,
giving $B=2$ and two copies of each load $0,1/2,1$.
The four bottom triples, each of weight $1/4$, complete the
fractional coverage of the state items and the single filler item. The first two bottom triples use $u=1/2$, and the last two use $u=1$.
Numbers above bins indicate weights; fractions below items indicate sizes.
Here $\varepsilon=1/10$, so $\lambda=1/100$ in~\eqref{eq:state-sizes}.
}
			\label{fig:state-copies}
		\end{figure}
		
		\Needspace{6\baselineskip}
		\begin{claim}\label{cl:universal-fractional}
			$L\le B<L+3/2$, $s_I(I)\le L$, $1\le B\le n+1$, and $m\le B(B+1)$.
			The weights $x$ satisfy~\eqref{eq:universal-fractional} and are computable in polynomial time.
		\end{claim}
		\begin{claimproof}
			Singletons give $L\le n$. Any feasible LP solution satisfies
			$s_I(I)\le\sum_U y_U s_I(U)\le\sum_U y_U$, so
			\begin{equation}\label{eq:universal-B}
				L\le B<L+3/2,\qquad s_I(I)\le L,\qquad 1\le B\le n+1.
			\end{equation}
			Since $m/(B+1)\le s_I(R)\le B$, we have $m\le B(B+1)$.
			The projected weights cover every item of $R$ at least once.
			At each step, $e_i\ge0$ and~\eqref{eq:universal-trim} gives
			$0\le\delta_C\le y_C$ and $\sum_{j=1}^a\delta_{C_j}=e_i$.
			Equation~\eqref{eq:universal-update} reduces the coverage of $i$ to one,
			preserves every other item's coverage and the total weight, and
			increases the number of positive coordinates by at most one: only one weight can be split
			between $C$ and $C\setminus\{i\}$ with both parts positive.
			Thus~\eqref{eq:universal-exact-lp} is feasible with cost at most $W$.
			It has $m$ equality constraints, so a basic solution uses at most
			$m$ nonempty configurations. Adding $x_\varnothing\ge0$ gives
			\begin{equation}\label{eq:universal-fractional}
				\sum_{\substack{C\in\mathcal C\\i\in C}}x_C=1\ (i\in R),\qquad
				\sum_{C\in\mathcal C}x_C=B,\qquad
				|\mathcal C\setminus\{\varnothing\}|\le m.
			\end{equation}
			All updates and the explicit LP have polynomial size and rational
			coefficients of polynomial binary length.
		\end{claimproof}
		
		\begin{claim}\label{cl:universal-states}
			$|V|\ge m+B$, $M=O((L+1)^4)$, and a feasible weight family exists
			for $(R,B,V,h)$; hence $\LP(J)=M$.
		\end{claim}
		\begin{claimproof}
			Positive item sizes make each configuration's partial loads distinct.
			By~\eqref{eq:universal-fractional}, $|H|\le m^2+1$ and
			\[
			\sum_{u\in H}\mu_u
			=\sum_{C\in\mathcal C}x_C(|C|+1)=m+B.
			\]
			Equation~\eqref{eq:universal-states} therefore gives
			\begin{equation}\label{eq:universal-size}
				m+B\le|V|\le m+B+|H|.
			\end{equation}
			Together with $m\le B(B+1)$ and~\eqref{eq:universal-B}, bound~\eqref{eq:universal-size}
			gives $M=O((L+1)^4)$.
			For each $C\in\mathcal C$ with its fixed order $(i_1,\ldots,i_k)$,
			and each choice $1\le r_j\le\lceil\mu_{h_j(C)}\rceil$ for $0\le j\le k$,
			give the path
			\[
			\pi=((h_0(C),r_0),i_1,(h_1(C),r_1),\ldots,i_k,(h_k(C),r_k))
			\]
			weight $x_C/\prod_{j=0}^k\lceil\mu_{h_j(C)}\rceil$; give all other paths
			weight zero. Consecutive loads differ by the corresponding item size,
			so these are state paths. Equation~\eqref{eq:universal-fractional}
			gives total weight $B$ and item coverage one. The total weight of paths
			containing $(u,r)$ is $\mu_u/\lceil\mu_u\rceil\le1$.
			Thus~\eqref{eq:state-fractional} holds, and Lemma~\ref{lem:states}(i)
			gives $\LP(J)=M$. These paths are a certificate and need not be enumerated.
		\end{claimproof}
		
		\begin{claim}\label{cl:universal-reinsert}
			Any packing of $R$ into $q$ bins extends in polynomial time to a
			packing of $I$ into at most $\max\{q,B+1\}$ bins.
		\end{claim}
		\begin{claimproof}
			Enumerate $I\setminus R=\{i_1,\ldots,i_z\}$ and the given packing
			as $(B_1^0,\ldots,B_q^0)$, with $q_0=q$. For $1\le r\le z$, set
			\[
			\begin{aligned}
				j_r&=\min\bigl(\{j:1\le j\le q_{r-1},\
				s_I(B_j^{r-1})+s_I(i_r)\le1\}\cup\{q_{r-1}+1\}\bigr),\\
				q_r&=\max\{q_{r-1},j_r\}.
			\end{aligned}
			\]
			If $j_r=q_{r-1}+1$, set $B_{j_r}^{r-1}=\varnothing$. Then set
			\[
			B_j^r=\begin{cases}
				B_j^{r-1}\cup\{i_r\},&j=j_r,\\
				B_j^{r-1},&j\ne j_r
			\end{cases}\qquad(1\le j\le q_r).
			\]
			If the final bin count $K=q_z$ exceeds $q$, every preceding bin had
			load greater than $B/(B+1)$ when the last bin opened, because
			$s_I(i_r)<1/(B+1)$. Thus
			$(K-1)B/(B+1)<s_I(I)\le B$, giving $K\le B+1$.
		\end{claimproof}
		
		By Lemma~\ref{lem:states}(iii), a packing of $J$ into $M+t$ bins gives
		a packing of $R$ into at most $B+96t$ bins. Claim~\ref{cl:universal-reinsert}
		extends it to a packing of $I$ into at most
		\[
		\max\{B+96t,B+1\}\le B+1+96t<L+3+96t.
		\]
		This proves Lemma~\ref{lem:universal-decode}; taking
		$t=\OPT(J)-M$ proves Theorem~\ref{thm:universal}.
		Claims~\ref{cl:universal-fractional}--\ref{cl:universal-states} give
		polynomially many states, each load a sum of input rational sizes.
		Hence the reduction and decoder run in polynomial time.
	\end{proof}
	
	By Lemma~\ref{lem:universal-decode}, for any function $g:[1,\infty)\to[0,\infty)$ and fixed $\varepsilon>0$,
	a polynomial-time algorithm using at most $\LP(J)+g(\LP(J))$ bins
	on every instance with sizes in
	$(1/3-\varepsilon,1/3+\varepsilon)\cap(1/4,1/2)$ therefore
	gives a polynomial-time algorithm using at most
	$\LP(I)+3+96g(M)$ bins on arbitrary instances, where
	$M=O((\LP(I)+1)^4)$.
	
	\section{Quantitative bounds for the five-item construction}\label{sec:source}
In this section, we bound and optimize the construction of~\cite{OAI26} to yield an $\Omega(\ls \OPT)$ additive gap. 
	For a graph $G=(V,E)$, a \emph{vertex cover} is a set
	$U\subseteq V$ meeting every edge, and $\tau(G)$ denotes its minimum
	cardinality. Let $K_4$ be the complete graph on four vertices.
	We use the following properties of the construction
	in~\cite{OAI26} and bound its output size below.
	
	\begin{proposition}[\cite{OAI26} Theorem~3.1 and Propositions~4.1, 5.1, 6.1]
		\label{prop:source}
		Fix $\rho=1/96$. For every integer $c\ge0$, simple graph
		$G=(V,E)$ with $n=|V|\ge1$ and $m=|E|\ge1$, and integer
		$k\in\{0,\ldots,n\}$, the algorithm of~\textnormal{\cite{OAI26}}, denoted $\mathcal A$, on input
		$(G,k,c)$ produces an integer $N\ge1$ and a bin packing instance $I$ with
		a partition $I=I_1\mathbin{\dot\cup}\cdots\mathbin{\dot\cup}I_5$,
		where $|I_j|=N$ for $1\le j\le5$ and all sizes lie in $(1/6,1)$.
		Let $\mathcal F=\{U\in\mathcal H(I):|U\cap I_j|=1\ \forall j\in\{1,\ldots,5\}\}$.
		Then:
		\begin{enumerate}
			\item If $\tau(G)\le k$, then $I$ has a partition into members of $\mathcal F$.
			\item If $\OPT(I)\le N+c$, then $\tau(G)\le k+n/96$.
			\item If $G=K_4$ and $k=2$, there are weights $(x_U)_{U\in\mathcal F}$
			with $x_U\ge0$ for all $U\in\mathcal F$ and
			$\sum_{U\in\mathcal F:\,i\in U}x_U=1$ for every $i\in I$.
		\end{enumerate}
		For every fixed $c$, the construction takes polynomial time.
	\end{proposition}
	
	We show that the construction remains polynomial-time for
	a suitable $c=\Theta(\ls n)$.
	To bound the output size when $c$ grows, we make the parameters of
	the source construction explicit; $d$ determines the number of
	tree-growth phases. Set $\rho=1/96$. For the following parameter formulas, fix
	an arbitrary input $(G,k,c)$ of $\mathcal A$ and define
	\begin{equation}\label{eq:source-parameters}
		\begin{aligned}
			P_0&=100^{11},&K&=25(c+1)(1+5P_0),\\
			A_0&=288000(1+5P_0),&d&=A_0(c+1)=11520K.
		\end{aligned}
	\end{equation}
	Here $P_0,A_0$ are absolute constants. These choices satisfy
	the source requirements $d\ge72K$ and $d\ge120K/\rho$.
	We first bound the trees in terms of $d$, then bound the item count
	and encoding length in terms of the trees, and finally choose
	$c=\Theta(\ls n)$ so that both bounds are polynomial in $n$.
	\subsection{Tree size}

	For a finite rooted tree $T$, let $V(T)$ be its vertex set, $r(T)$ its
	root, $V_\ell(T)$ its vertices at distance $\ell$ from the root, and
	$\Pi(T)$ the family of vertex sets of its root-to-leaf paths.
	A marking is a subset of $V(T)\setminus\{r(T)\}$; it hits a path
	if their intersection is nonempty.
	
	\paragraph{Admissible trees.}
	Fix an integer $d\ge1$. A \emph{$d$-admissible tree} is a finite rooted
	tree $T^\circ$ with disjoint red and blue markings satisfying the following
	conditions. All leaves have depth $L\ge1$, every root-to-leaf path has
	exactly $2d-1$ marked vertices, and each depth $1\le\ell\le L$ has
	the same positive number $p_\ell$ of red and blue vertices.
	In separate copies, retain the union of root-to-leaf paths with at least
	$d$ red (respectively, blue) vertices to obtain $T^+$ (respectively, $T^-$),
	without contracting edges; both copies must be nonempty.
	For $\sigma\in\{+,-\}$, label each retained vertex $(\sigma,v)$ by its
	original $v\in V(T^\circ)$ and set $\varphi^\sigma(\sigma,v)=v$.
	Let $\mathcal M^+$ and $\mathcal M^-$ be the retained red and blue vertices,
	respectively. Both copies have leaf depth $L$, and for $\sigma\in\{+,-\}$
	we have
	\[
	\begin{aligned}
		|\mathcal M^\sigma\cap V_\ell(T^\sigma)|&\le p_\ell
		&&(1\le\ell\le L),\\
		|\mathcal M^\sigma\cap\pi|&\ge d
		&&(\pi\in\Pi(T^\sigma)).
	\end{aligned}
	\]
	Finally, any markings $\mathcal A^+,\mathcal A^-$ of $T^+,T^-$ satisfying
	\[
	|\mathcal A^+\cap V_\ell(T^+)|+|\mathcal A^-\cap V_\ell(T^-)|\le p_\ell
	\quad(1\le\ell\le L)
	\]
	must leave a path unhit: $\mathcal A^\sigma\cap\pi=\varnothing$ for some
	$\sigma\in\{+,-\}$ and $\pi\in\Pi(T^\sigma)$.
	
	Set
	\[
	P=\sum_{\ell=1}^L p_\ell,\qquad H=|V(T^+)|+|V(T^-)|,
	\]
	and let $D_*\ge1$ be an integer upper bound on every vertex's number
	of children in either copy. Define the tower sequence
	\begin{equation}\label{eq:towers}
		T_0=2,\qquad T_{j+1}=2^{T_j}\quad(j\ge0).
	\end{equation}
	\Needspace{5\baselineskip}
	The next lemma bounds each phase of the competing-tree construction
	of~\cite[Section~2]{OAI26} by two additional exponentiations.
	
	\begin{lemma}\label{lem:tree}
		Given an integer $d\ge1$, one can deterministically construct a
		$d$-admissible tree, its marked copies, and integers $p_1,\ldots,p_L,D_*,S_d$
		in time $S_d^{O(1)}$, where
		\[
		\max\{d,L,P,D_*,H\}\le S_d\le T_{4d+2}.
		\]
	\end{lemma}
	\begin{proof}
		Set $q=2d-1$. Start with one unmarked root, $Z_0=1$, and $L_0=0$.
		For $0\le a<q$, number the current leaves $1,\ldots,Z_a$ in creation
		order and perform the following operations for each leaf $i$:
		\begin{enumerate}
			\item Attach $2^i$ paths of length $i$, numbered $1,\ldots,2^i$.
			Mark the first $2^{i-1}$ endpoints red and the others blue.
			\item If $i<Z_a$, attach to each new endpoint
			$1+\sum_{j=i+1}^{Z_a}2^{j-1}$ paths of length $Z_a-i$.
			If $i=Z_a$, add no further edges.
		\end{enumerate}
		Every attached path uses fresh vertices except for its starting
		vertex; all new vertices other than the specified colored endpoints
		are unmarked. Let $Z_{a+1}$ be the resulting number of leaves, and set
		\begin{equation}\label{eq:tree-phase}
			L_{a+1}=L_a+Z_a,\qquad p_{L_a+i}=2^{i-1}\quad(1\le i\le Z_a).
		\end{equation}
		Let $T^\circ$ be the final tree and form $T^+,T^-$ and their markings
		as in the definition of admissibility. Set $L=L_q$, $Z=Z_q$,
		$D_*=Z$, and $S_d=10Z^2$.
		Figure~\ref{fig:tree-growth} shows the first two phases.
		\begin{figure}[htbp]
			\centering
			\begin{tikzpicture}[font=\small,x=1cm,y=.75cm]
				\node[bpvertex] at (0,0) {};
				\node at (0,-2.85) {$Z_0=1$};
				\draw[bpedge] (2.4,0)--(1.9,-1);
				\draw[bpedge] (2.4,0)--(2.9,-1);
				\node[bpvertex] at (2.4,0) {};
				\node[bpredmark] at (1.9,-1) {};
				\node[bpbluemark] at (2.9,-1) {};
				\node at (2.4,-2.85) {$Z_1=2$};
				\begin{scope}[shift={(7,0)},x=.56cm,y=.5cm]
					\draw[bpedge] (0,0)--(-2.7,-1)--(-4.2,-2);
					\draw[bpedge] (-2.7,-1)--(-1.2,-2);
					\draw[bpedge] (0,0)--(2.4,-1);
					\foreach \x in {-5.1,-4.2,-3.3} {\draw[bpedge] (-4.2,-2)--(\x,-3);}
					\foreach \x in {-2.1,-1.2,-.3} {\draw[bpedge] (-1.2,-2)--(\x,-3);}
					\foreach \x in {1.05,1.95,2.85,3.75} {
						\draw[bpedge] (2.4,-1)--(\x,-2)--(\x,-3);
						\node[bpvertex] at (\x,-2) {};
					}
					\node[bpvertex] at (0,0) {};
					\node[bpredmark] at (-2.7,-1) {};
					\node[bpbluemark] at (2.4,-1) {};
					\node[bpredmark] at (-4.2,-2) {};
					\node[bpbluemark] at (-1.2,-2) {};
					\foreach \x in {-5.1,-4.2,-3.3,-2.1,-1.2,-.3} {\node[bpvertex] at (\x,-3) {};}
					\foreach \x in {1.05,1.95} {\node[bpredmark] at (\x,-3) {};}
					\foreach \x in {2.85,3.75} {\node[bpbluemark] at (\x,-3) {};}
				\end{scope}
				\node at (6.65,-2.85) {$Z_2=10$};
			\end{tikzpicture}
			\caption{The unpruned tree before any growth and after its first two
				phases. Red triangles and blue squares are marked vertices; gray
				circles are unmarked. All leaves in each panel have a common depth.
				The leaf counts are $Z_0=1$, $Z_1=2$, and $Z_2=10$; the next phase
				has $Z_3=699050$ leaves.}
			\label{fig:tree-growth}
		\end{figure}
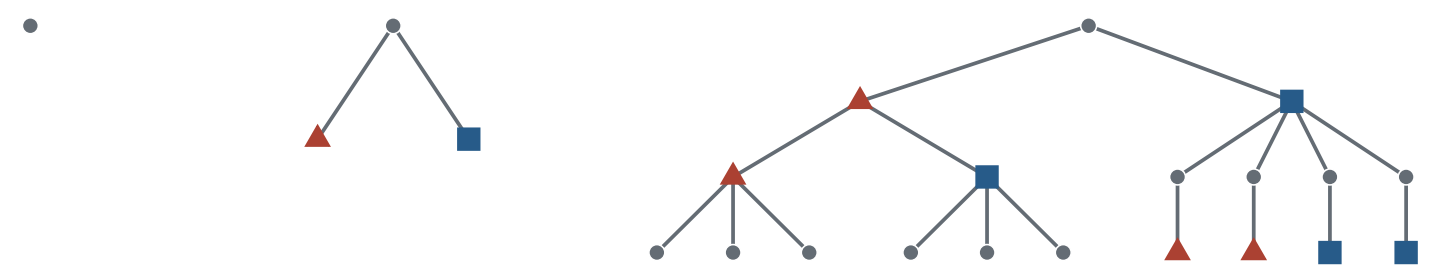
		
		\begin{claim}\label{cl:tree-counts}
			All final leaves have depth $L$, every root-to-leaf path has $q$ marks,
			and depth $\ell$ has $p_\ell$ marks of each color.
			Also, $2Z_a\le Z_{a+1}\le2^{2Z_a}$ for $0\le a<q$.
		\end{claim}
		\begin{claimproof}
			At phase $a$, each old leaf gains paths of total length $Z_a$,
			each with one colored endpoint. Only leaf $i$ creates marks at depth
			$L_a+i$, exactly $2^{i-1}$ of each color. Induction and~\eqref{eq:tree-phase}
			give the depth and marking assertions. Counting the new leaves gives
			\begin{equation}\label{eq:leaf-recurrence}
				\begin{aligned}
					Z_{a+1}&=\sum_{i=1}^{Z_a}2^i
					\left(1+\sum_{j=i+1}^{Z_a}2^{j-1}\right)\\
					&=\sum_{i=1}^{Z_a}2^i(1+2^{Z_a}-2^i)
					=\frac23(4^{Z_a}-1).
				\end{aligned}
			\end{equation}
			This lies between $2Z_a$ and $2^{2Z_a}$.
		\end{claimproof}
		
		\begin{claim}\label{cl:tree-admissible}
			$T^\circ$ is $d$-admissible.
		\end{claim}
		\begin{claimproof}
			An all-red path and an all-blue path exist, so both copies are nonempty.
			Each copy is a union of paths to depth $L$, with at least $d$ marks
			of its color per path and at most $p_\ell$ at depth $\ell$, by
			Claim~\ref{cl:tree-counts}.
			For the shared-quota condition, first fix a marking $\mathcal A$
			of $T^\circ$ satisfying $|\mathcal A\cap V_\ell(T^\circ)|\le p_\ell$
			for $1\le\ell\le L$.
			Suppose an unhit path reaches leaf $i$ at the start of phase $a$.
			By~\eqref{eq:tree-phase},
			\[
			\sum_{j=1}^i p_{L_a+j}=2^i-1<2^i.
			\]
			Thus one of the $2^i$ initial paths has no mark of $\mathcal A$.
			If $i<Z_a$, its endpoint has
			$1+\sum_{j=i+1}^{Z_a}p_{L_a+j}$ continuations, disjoint except at their
			common endpoint outside $\mathcal A$. At most $\sum_{j=i+1}^{Z_a}p_{L_a+j}$ of these
			can meet $\mathcal A$, so one is unhit. Starting at the unmarked root
			and repeating this argument gives an unhit path in $\Pi(T^\circ)$.
			Now let $\mathcal A^+,\mathcal A^-$ satisfy the shared quotas and set
			$\mathcal A=\varphi^+(\mathcal A^+)\cup\varphi^-(\mathcal A^-)$.
			The depth-preserving projections give
			\[
			|\mathcal A\cap V_\ell(T^\circ)|
			\le|\mathcal A^+\cap V_\ell(T^+)|+|\mathcal A^-\cap V_\ell(T^-)|
			\le p_\ell.
			\]
			An unhit path in $T^\circ$ has $2d-1$ colored vertices, hence at least
			$d$ of one color. Its copy in the corresponding tree is unhit by
			that tree's marking.
		\end{claimproof}
		
		\Needspace{6\baselineskip}
		\begin{claim}\label{cl:tree-size}
			$\max\{d,L,P,D_*,H\}\le S_d\le T_{4d+2}$, and the construction
			takes time $S_d^{O(1)}$.
		\end{claim}
		\begin{claimproof}
			The doubling bound of Claim~\ref{cl:tree-counts} gives
			\[
			L=\sum_{a=0}^{q-1}Z_a\le Z,\qquad
			H\le2(L+1)Z.
			\]
			A vertex at a phase with $Z_a$ leaves has at most
			$2^{Z_a}\le Z_{a+1}\le Z=D_*$ children; pruning cannot increase this.
			Equation~\eqref{eq:leaf-recurrence} also gives
			\[
			P=\sum_{a=0}^{q-1}(2^{Z_a}-1)
			\le\sum_{a=1}^q Z_a\le2Z.
			\]
			Since $d\le Z$, all five parameters are at most $S_d=10Z^2$.
			Using~\eqref{eq:towers}, $2x\le2^x$ for integers $x\ge2$, and $Z_0=1\le T_0$,
			induction yields
			\[
			Z_{a+1}\le2^{2Z_a}\le2^{2T_{2a}}
			\le2^{2^{T_{2a}}}=T_{2a+2}\quad(0\le a<q).
			\]
			For every integer $j\ge2$, $10T_j^2\le T_{j+2}$. Therefore
			$S_d\le10T_{4d-2}^2\le T_{4d}\le T_{4d+2}$.
			The unpruned tree has at most $(L+1)Z\le2Z^2$ vertices.
			Traversing it constructs both copies and lists their marks and
			$p_1,\ldots,p_L$. All computed integers have $O(\log S_d)$ bits,
			so the running time is polynomial in $S_d$.
		\end{claimproof}
		Claims~\ref{cl:tree-admissible} and~\ref{cl:tree-size} prove the lemma.
	\end{proof}
	
	\subsection{Item counts and encoding lengths}
	We now bound the number of items and encoding bits used and prove polynomial-time generation for a growing gap.
	For this subsection, take any input $(G,k,c)$ of $\mathcal A$, with
	$G=(V,E)$, $n=|V|$, and $m=|E|$. Set $K,d$ as in
	\eqref{eq:source-parameters}. Apply Lemma~\ref{lem:tree} to this $d$
	and use the resulting $L,p_1,\ldots,p_L,P,H,D_*$.
	We run $\mathcal A$ with these trees and our choices of $d,D_*$.
We use the repetition count $R$ of~\cite[Section~3.1]{OAI26}.
The item count follows from the tree-row and job inventories
in~\cite[Section~3.2, Eqs.~(3.4) and (3.6)]{OAI26}.
The common denominator follows from the scalar encoding
and coordinate denominators in~\cite[Eqs.~(3.8) and (3.9)]{OAI26}:
	\begin{align}
		R&=100(P+nK+1),&
		N&=n(dH+P)+4dmR,\label{eq:item-count}\\
		D_{\rm size}
		&=160\cdot10^9P_0\,3^n(R+1)^m
		[100(D_*+1)]^L.\label{eq:denominator}
	\end{align}
	Every item size is a positive integer multiple of
	$1/D_{\rm size}$ and is smaller than one.
	By~\eqref{eq:denominator}, its binary length is
	\begin{equation}\label{eq:bits}
		O\bigl(n+m\log(R+1)+L\log(D_*+1)\bigr).
	\end{equation}
	\Needspace{5\baselineskip}
	Choosing the additive gap to be a sufficiently small multiple
	of $\ls n$ keeps the trees smaller than the input graph.
	\begin{lemma}\label{lem:uniform}
		There is a computable integer function $c(n)=\Theta(\ls n)$,
		defined for all sufficiently large $n$, such that for every simple
		graph $G$ with $n$ vertices and at least one edge and every integer
		$0\le k\le n$, the algorithm $\mathcal A$ of
		Proposition~\ref{prop:source}, on input $(G,k,c(n))$ with parameters
		\eqref{eq:source-parameters}, runs in uniform deterministic polynomial time.
		Its output has $N=O(n^5)$ items per class, each with $O(n^2\log n)$ bits.
	\end{lemma}
	\begin{proof}
		We choose $c(n)$ so that the tree bound is smaller than $n$.
		Let $h=\ls n$ and let $A_0$ be as in~\eqref{eq:source-parameters}.
		Define
		\begin{equation}\label{eq:allowance}
			c(n)=\left\lfloor\frac{h-4}{4A_0}\right\rfloor-1,
			\qquad d=A_0(c(n)+1).
		\end{equation}
		For sufficiently large $n$, \eqref{eq:allowance} gives
		$c(n)\ge0$, $c(n)=\Theta(h)$, and $4d+2\le h-2$.
		The definition of $\ls$ gives $T_{h-2}<n\le T_{h-1}$ for $h\ge2$.
		Thus, Lemma~\ref{lem:tree} implies
		\[
		S_d\le T_{4d+2}\le T_{h-2}<n.
		\]
		To compute $h$ for $n\ge2$, start at $T_0=2$ and successively
		compute $\min\{n,2^{T_j}\}$ until reaching $n$ at index $h-1$.
		Test $T_j\ge\lceil\log_2 n\rceil$ before exponentiating, so every
		stored integer has $O(\log n)$ bits. This computes $h$ and $c(n)$
		in polynomial time.
		
		We now bound the source instance. By Lemma~\ref{lem:tree} and
		\eqref{eq:source-parameters}, we have $d,K,L,P,D_*,H\le n$. Since $m\le n^2$,
		\eqref{eq:item-count} gives $R=O(n^2)$ and $N=O(n^5)$.
		Equation~\eqref{eq:bits} gives $O(n^2\log n)$ bits per item.
		The trees have at most $S_d<n$ vertices; listing the $5N$
		items and computing their rational sizes uses polynomially
		many operations on polynomial-bit integers. Hence, the full
		construction runs in polynomial time, with an exponent
		independent of $c(n)$.
	\end{proof}
	
	\section{Proofs of the main theorems}\label{sec:applications}
	We now combine the reductions to prove the additive hardness and configuration-LP gap lower bounds.
	We use the elementary bound
	\begin{equation}\label{eq:logstar-power}
		\ls(ax^b)\le\ls x+O(1)\qquad(x\to\infty)
	\end{equation}
	for fixed $a,b>0$. For all sufficiently large $x$, we have
	$ax^b\le2^x$, and therefore
	$\ls(ax^b)\le\ls(2^x)=1+\ls x$.
	
	\Needspace{8\baselineskip}
	\subsection{Additive hardness}
We prove that polynomial-time bin packing cannot achieve additive error $o(\ls\OPT)$ unless $\mathrm P=\mathrm{NP}$.
	The following construction converts a satisfiability gap into
	a vertex-cover gap while preserving the exact YES-case optimum.
	\begin{lemma}\label{lem:vc}
		For simple graphs $G$ with $n=|V(G)|$ divisible by three, it is
		NP-hard to distinguish the following two cases:
		\begin{enumerate}
			\item $\tau(G)=2n/3$.
			\item $\tau(G)\ge2n/3+n/48$.
		\end{enumerate}
	\end{lemma}
	\begin{proof}
		Consider Boolean formulas $\Phi=\bigwedge_{r=1}^q C_r$ on variables
		$z_1,\ldots,z_p$, where $p,q\ge1$ are integers and each clause
		$C_r=\ell_{r,1}\vee\ell_{r,2}\vee\ell_{r,3}$, $1\le r\le q$, has
		three literal occurrences, each equal to $z_j$ or $\neg z_j$ for
		some $j\in\{1,\ldots,p\}$.
		Let
		\[
		s(\Phi)=\max_{a\in\{0,1\}^p}
		|\{r\in\{1,\ldots,q\}:C_r(a)=1\}|.
		\]
		H\aa stad's gap theorem~\cite[Theorem~6.5]{Has01} states that it is
		NP-hard to distinguish $s(\Phi)=q$ from $s(\Phi)\le15q/16$.
		Write $\overline{z_j}=\neg z_j$ and
		$\overline{\neg z_j}=z_j$ for $1\le j\le p$.
		The graph has clause triangles and edges between complementary literals:
		set $G=(V,E)$ with
		\[
		\begin{aligned}
			V&=\{(r,j):1\le r\le q,\ 1\le j\le3\},\\
			E&=\bigl\{\{(r,j),(s,k)\}\subseteq V:\ (r,j)\ne(s,k),\
			(r=s)\lor(\ell_{r,j}=\overline{\ell_{s,k}})\bigr\}.
		\end{aligned}
		\]
		Thus, $G$ is simple and has $n=3q$ vertices.
		
		An independent set of $G$ contains at most one occurrence from
		each clause and contains no complementary pair. Setting its
		chosen literals true satisfies at least as many clauses as its
		cardinality. Conversely, choosing one true literal from every
		satisfied clause gives an independent set. Thus, the maximum independent-set size of $G$ is $s(\Phi)$, and
		$\tau(G)=3q-s(\Phi)$. In the YES case $s(\Phi)=q$, giving
		$\tau(G)=2q=2n/3$. In the NO case $s(\Phi)\le15q/16$, giving
		$\tau(G)\ge2q+q/16=2n/3+n/48$.
		Taking a fixed number of disjoint copies preserves both
		ratios and makes $n$ exceed any prescribed constant.
	\end{proof}
	
	\begin{proof}[Proof of Theorem~\ref{thm:hardness}]
		We compose the two reductions. Fix a rational $\varepsilon>0$.
		Lemmas~\ref{lem:uniform} and~\ref{lem:reduction} give absolute constants
		$a,D>0$ and an integer $n_0\ge1$ such that, for every $n\ge n_0$,
		$c(n)\ge a\ls n$ and applying Lemma~\ref{lem:reduction} to the output
		of $\mathcal A(G,k,c(n))$ for any admissible $G,k$ with $|V(G)|=n$
		produces $3M$ items with $M\le Dn^{20}$.
		Choose these constants so that also
		$\ls(Dn^{20})\le2\ls n$ for all $n\ge n_0$, as permitted by
		\eqref{eq:logstar-power}, and set $\alpha=a/192$.
		Take a graph $G$ from Lemma~\ref{lem:vc} with $n=|V(G)|\ge n_0$,
		using the disjoint-copy padding in its proof, and set
		$k=2n/3$ and $c=c(n)$.
		Apply Proposition~\ref{prop:source} to $G,k,c$, obtaining
		an instance $I$ with $N$ items in each of its five classes.
		Apply Lemma~\ref{lem:reduction} to $I$, obtaining $J$ with
		$3M$ items, where $M=O(N^4)$.
		
		In the YES case $\tau(G)=k$, so
		Proposition~\ref{prop:source}(i) and
		Lemma~\ref{lem:reduction}(i) give $\OPT(J)=M$.
		In the NO case $\tau(G)\ge k+n/48>k+n/96$,
		so Proposition~\ref{prop:source}(ii) implies
		$\OPT(I)>N+c$.
		Write $\OPT(J)=M+t$ with integer $t\ge0$; this is possible
		because $|J|=3M$ and each bin holds at most three items.
		By Lemma~\ref{lem:reduction}(iii), we have
		$\OPT(I)\le N+C_0t$, where $C_0=96$. Hence, we have
		\[
		\OPT(J)=M+t>M+c/C_0.
		\]
		
		We express the additive gap in terms of $M$.
		By the choices of $a,D,n_0$ and $\alpha$, we have
		\[
		\frac{c(n)}{96}\ge\frac{a}{96}\ls n
		\ge\frac{a}{192}\ls M=\alpha\ls M.
		\]
		The reduction is polynomial-time, so the displayed YES and
		NO cases establish the stated NP-hardness.
		
		For the approximation consequence, suppose a polynomial-time
		algorithm uses at most $\OPT(J)+g(\OPT(J))$ bins, where
		$g:[1,\infty)\to[0,\infty)$ satisfies $g(x)=o(\ls x)$. As $M\ge N\ge n$ tends to infinity, on sufficiently large
		YES instances the algorithm produces fewer than
		$M+\alpha\ls M$ bins, whereas this is impossible on NO
		instances. This decides the NP-hard promise problem in
		polynomial time.
	\end{proof}
	
	\subsection{Configuration-LP gap}
For the following proof, we construct three-item instances with unconditional configuration LP gap $\Omega(\ls\OPT)$.
	\begin{proof}[Proof of Theorem~\ref{thm:gap}]
		We construct an instance $J$ with $\LP(J)=M$ and
		$\OPT(J)>M+c/96$ by applying the five-item construction to $K_4$
		and then reducing its output to three-item bins.
		Fix a rational $\varepsilon>0$ and an integer $c\ge1$.
		Apply Proposition~\ref{prop:source} with $G=K_4$, $k=2$
		and the parameters~\eqref{eq:source-parameters}. Let $I$
		be the resulting instance, with $N$ items per class.
		Since $\tau(K_4)=3>2+4/96$,
		Proposition~\ref{prop:source}(ii) implies
		$\OPT(I)>N+c$. Proposition~\ref{prop:source}(iii)
		gives a fractional partition of $I$ into feasible five-tuples.
		Applying Lemma~\ref{lem:reduction} yields an instance
		$J$ with $3M$ items, where $M=O(N^4)$, satisfying
		\begin{equation}\label{eq:gap-transfer}
			\begin{aligned}
				\LP(J)&=M,\\
				\OPT(J)&>M+c/C_0,\\
				C_0&=96.
			\end{aligned}
		\end{equation}
		
		We bound $\ls\OPT(J)$ in terms of $c$.
		Here $n=4$ and $m=6$. Let $S_d$ be the bound from
		Lemma~\ref{lem:tree} for $d=A_0(c+1)$. Then~\eqref{eq:item-count} gives
		$N=O(S_d^2)$ and $M=O(S_d^8)$.
		Every pair of target items fits, so we have
		\[
		M\le\OPT(J)\le\lceil3M/2\rceil=O(S_d^8).
		\]
		Lemma~\ref{lem:tree}, \eqref{eq:source-parameters}, and
		\eqref{eq:logstar-power} imply
		\[
		\ls\OPT(J)\le\ls S_d+O(1)
		\le4d+O(1)=O(c+1).
		\]
		Choose an absolute $C\ge1$ such that
		$\ls\OPT(J)\le C(c+1)$ for all $c\ge1$.
		Since $c+1\le2c$, \eqref{eq:gap-transfer} gives
		\[
		\OPT(J)-\LP(J)>\frac c{C_0}
		\ge\frac{\ls\OPT(J)}{2CC_0}.
		\]
		Take $\beta=1/(2CC_0)$. Finally,
		$M+c/C_0<\OPT(J)\le\lceil3M/2\rceil\le3M/2+1$ gives
		$M>2c/C_0-2$, so the integers $M$ are arbitrarily large.
	\end{proof}

	\section{Discussion}
	The results give additive configuration LP gaps and approximation
	hardness of order $\Omega(\ls\OPT)$, even with all sizes arbitrarily
	close to $1/3$.
	More broadly, Theorem~\ref{thm:universal} further shows that this restricted class
	captures the additive LP gap of arbitrary instances up to a constant
	factor and an additive constant, with polynomial growth in the LP value, confirming the conjecture that these 3-partition instances are, at least in the sense of integrality gap of the canonical configuration LP and up to constants, the hardest bin packing instances. We believe this result may be useful in other settings. 
	
	The true asymptotic additive integrality gap and, assuming
	$\mathrm P\ne\mathrm{NP}$, the optimal additive polynomial-time
	approximation guarantee remain open. Our
	$\Omega(\ls\OPT)$ lower bounds leave a wide gap to the
	$O(\log\OPT)$ upper bounds of~\cite{HR17}.
	It remains to determine whether either lower bound can be raised to
	$\Omega(\log\OPT)$, whether the upper bounds can be reduced,
	and whether the integrality gap and computational hardness have the
	same asymptotic order.

\section*{Acknowledgments}
	Ilan Doron-Arad is supported by grant NSF DMS-2031883 and Vannevar Bush Faculty Fellowship ONR-N00014-20-1-2826 (PI Mossel).
    
	\paragraph{AI methodology.}
	The author used ChatGPT (OpenAI) in iterative interactions to develop
	the results and proofs in Sections~\ref{sec:reduction}--\ref{sec:applications}
	and to draft and revise the manuscript, including the proof overview
	and figures. This use materially influenced the technical arguments
	and their presentation. The author is responsible for the accuracy,
	originality, and integrity of all content, including AI-assisted
	material and references.

\end{document}